\documentclass[superscriptaddress,amssymb,nofootinbib,onecolumn,aps,prd,longbibliography]{revtex4-2}
\usepackage{amsmath}
\usepackage{hyperref}
\hypersetup{colorlinks=true, urlcolor=blue,linkcolor=blue, citecolor=blue}
\usepackage[utf8]{inputenc}
\usepackage{graphicx}
\usepackage{mathtools}%para usar floor
\usepackage{slashed}
\usepackage{longtable}

\DeclarePairedDelimiter\round{\lfloor}{\rceil} %para facilitar uso do floor

\begin{document}
	\title{Feynman Amplitudes in Periodically Compactified Spaces - Spin 0}
	\author{E. Cavalcanti}
	\email[]{erich@cbpf.br}
	\affiliation{Centro Brasileiro de Pesquisas F\'{\i}sicas/MCTI, 22290-180 Rio de Janeiro, RJ, Brazil}
	
	\begin{abstract}
		We propose an extension of the Schwinger parametric representation for Feynman amplitudes in $D$ euclidean dimensions to a scenario where $d$ dimensions are compactified ($d<D$) through the introduction of periodic boundary conditions in space. We obtain two valid representations, one useful near the bulk (large compactification length) and another useful near the dimensional reduction (small compactification length). Also, to illustrate, we exhibit some Feynman amplitudes up to three loops in a compactified scalar field theory.
	\end{abstract}
	
	\maketitle
	
	%%%%%%%%%%%%%%%%%%%%%%%%%%%%%%%%%%%%%%%%%%%%%%%%%%%%%%%%%%%%%%%
	%%%%%%%%%%%%%%%%%%%%%%%%%%%%%%%%%%%%%%%%%%%%%%%%%%%%%%%%%%%%%%%
	%%%%%%%%%%%%%%%%%%%%%%%%%%%%%%%%%%%%%%%%%%%%%%%%%%%%%%%%%%%%%%%
	
	\section{Introduction}
	One essential task in perturbative quantum field theory is to obtain, for a given theory, the amplitude of graph $G$ at some arbitrary order. This task is decades older and its solution is already textbook material, the most known parametric representations are those of Feynman~\cite{ZinnJustin:2002ru,Itzykson:1980rh}, Schwinger~\cite{ZinnJustin:2002ru,Itzykson:1980rh,Rivasseau:1991ub} and Mellin~\cite{Paris:2001book,Smirnov:2012gma} although other efforts, as for example the Complete Mellin representation~\cite{deCalan:1979ii,deCalan:1980aew,Malbouisson:1999td,Linhares:2006bi,Gurau:2007az}, were also studied.
	The employment of these representations in a general setting employs the Symanzik polynomials, which can be directly extracted from the topology of the graph by inspecting the 1-trees and 2-trees~\cite{Itzykson:1980rh,Rivasseau:1991ub,Gurau:2014vwa}. That way, one can avoid the otherwise cumbersome computation of these polynomials and we have an `easy' prescription to compute higher-order diagrams. This first simple task was essential for the important developments that follow, for example, proving the renormalizability of a quantum field theory up to all orders or extracting the asymptotic behavior of a diagram.
		
	Although this topic, to obtain a parametric representation, is completely settled for theories in non-compact space-time -- let us refer to the euclidean space-time with $D$ dimensions -- it has not yet been established for theories in compactified dimensions. 
	If the only interest of someone is the proof of renormalizability, one can justify this lack of understanding concerning theories in compactified dimensions due to the knowledge that the divergent behavior in the amplitudes of compactified theories come from the bulk, that is, the contribution related to the non-compactified space~\cite{Appelquist:1981vg,Landsman:1986uw}. This means that there is no need to prove renormalizability again as it comes directly from the proof in the bulk scenario. However, this does not justify the absence of this exploration in the literature. Right now, we do not have a prescription to `easily' compute higher-order diagrams in a scenario with compactified spaces neither an established asymptotic expansion for this scenario. The first step, if one is interested in some progress in topics that depend on higher-order corrections of Feynman diagrams in compactified spaces, is to establish a useful parametric representation and this is the purpose of this work.
	
	There are plenty of ways to introduce compactified dimensions, just as there are many possible choices of boundary conditions. Perhaps the most simple scenario are periodically compactified theories with just one compactified dimension, which is exactly the highly explored scenario of field theory at finite temperature~\cite{Landsman:1989be,LeBellac:1991cq,Das:1997gg,Kapusta:2006pm,Khanna:2009zz}, where the inverse temperature $\beta=1/T$ is introduced as the compactification length of a periodically compactified dimension throught the Matsubara formalism of imaginary time. In recent years, there is also a growing interest in field theory with a small circle compactification \cite{Tanizaki:2017qhf,Hetrick:1989vm,Aitken:2017ayq,Nitta:2015tua,Kanazawa:2017mgw}, which is somewhat equivalent to finite temperature but allows to impose antiperiodic or even twisted boundary conditions in the spatial compactification -- recall that the boundary condition in imaginary time is restricted, by the Kubo-Martin-Schwinger (KMS) condition~\cite{Landsman:1989be,LeBellac:1991cq,Das:1997gg,Kapusta:2006pm,Khanna:2009zz}, to follow the bosonic/fermionic nature of the field, but we are free of it if we deal with spatial compactification. When it comes to the extension to more compactified dimensions and different boundary conditions the range of applicability just grows even more: Camir effect~\cite{Plunien:1986ca,Bordag:2001qi,Milton:2004ya, Klimchitskaya:2009cw,Elizalde:2012zza}, superstring theory~\cite{Schwarz:1982jn,Horava:1996ma,Aharony:1999ti}, quantum field theory with an small extradimension~\cite{Chakraverty:2002qk,Agashe:2004ay,DaRold:2005mxj,Panico:2005dh,Panico:2006em,Blanke:2008yr,CcapaTtira:2010ez,Ospedal:2017ubh,Bhardwaj:2019fzv}, finite volume considerations QFT models and in particle physics~\cite{Gasser:1987zq,Lin:2001ek,Colangelo:2005gd,Aoki:2007ka,Hayakawa:2008an,Palhares:2009tf,Bernard:2010fp,Briceno:2012yi,Bhattacharyya:2012rp,Perez:2014sqa,Bhattacharyya:2014uxa,Bhattacharyya:2015kda,Juricic:2016tpt,Li:2017zny}, and so on. To see more applications we refer to the works~\cite{Elizalde:2012zza,Khanna:2009zz,Khanna:2014qqa,Mogliacci:2018oea} and references therein.
	
	In the last decades, Refs.~\cite{Khanna:2009zz,Khanna:2014qqa} established the so-called `quantum field theory in toroidal topologies', a formalism to deal with periodically compactificatified dimensions. This is born as an extension from the Matsubara prescription of imaginary time to more dimensions, producing a topology $\Gamma^d_D = \mathbb{R}^{D-d}\times\mathbb{S}_{L_1}\times\cdots\times\mathbb{S}_{L_d}$, where $\mathbb{S}_{L_i}$ represents each of the $d$ compactifications that can be imposed by periodic/antiperiodic boundary conditions~\footnote{One can easily extend it to quasiperiodic (also called twisted or anyonic) boundary conditions~\cite{Cavalcanti:2017wnm}}, with $L_i$ as the characteristic compactification length.
	Within this formulation, many applications were explored both in the field of particle physics and in the field of condensed matter, as the occurrence of size-dependent phase transitions~\cite{Abreu:2003zz,Abreu:2005uf,Linhares:2006my,Abreu:2009zz,Abreu:2011zzc,Abreu:2011rj,Linhares:2011nh,Khanna:2012zz,Khanna:2012js,Linhares:2012vr,Abreu:2013nca,Correa:2013mta,Cavalcanti:2017pjz}. In recent years, a somewhat similar approach~\cite{Mogliacci:2018oea} also started to explore the formulation and applications of periodically compactified QFT.
	
	So far, only few attempts were done~\cite{Benhamou:1990cd,Benhamou:1992,Khanna:2009tv,Linhares:toappear} in the direction to establish a parametric representation for Feynman amplitudes in compactified spaces. Three decades ago Benhamou~\cite{Benhamou:1990cd,Benhamou:1992} proposed a parametric representation for field theory at a finite temperature that recovers the Symanzik polynomials for the compactified part and that has as the zero-temperature limit the usual non-compactified parametric representation. One decade ago there was an attempt~\cite{Khanna:2009tv} to build for scalar field models some representation for the scenario of periodically compactified dimensions. Also, a recent work Ref.~\cite{Linhares:toappear} deals with the parametric representation for fields with different spins in a compactified space.
	
	In Sec.~\ref{Sec:ParRep} we start building the Schwinger parametric representation for a graph in periodically compactified spaces. We see that there are two possible paths to follow: 1 - we can consider the scale where the lengths of the compactified dimensions are very small (Sec.~\ref{Sec:NearRedDim}), meaning that we are close to a dimensional reduction; 2 - or we can assume that the lengths are very large (Sec.~\ref{Sec:NearBulk}), so we are close to the bulk scenario without compactifications. Both paths are equivalent and one could transport from one to another~\cite{Cavalcanti:2018pgi,Cavalcanti:2019mli}, but each one of them is more useful in one regime (near the dimensional reduction or the bulk), due to quicker convergence. After this, we illustrate the representation by showing some diagrams in Sec.~\ref{Sec:Diagrams} with the useful information to write their parametric representation. In the conclusions, we indicate some further developments.

	\section{Parametric representation of compactified Feynman diagrams}\label{Sec:ParRep}
	
	Let us start considering a general scalar scenario~\cite{ZinnJustin:2002ru,Itzykson:1980rh},
	\begin{equation}
	\widetilde{\mathcal{I}}_G = C_G \prod_{i=1}^I \left[\int \frac{d^D K_i}{(2\pi)^D}\frac{1}{K_i^2+m_i^2}\right]
	\prod_{v=1}^V \left[(2\pi)^D \delta^D \left(P_v-\sum_{i} \varepsilon_{vi} K_i\right)\right],
	\end{equation}
	\noindent where $D$ is the number of dimensions (we consider a euclidian space-time), $C_G$ is a factor related to the vertices and the symmetry of the graph, $K_i$ are the internal momenta -- that runs from $i=1,\ldots,I$, where $I$ is the number of internal lines --, $P_v$ are the sum of external momenta inciding over the vertex $v$ -- that runs from $v=1,\ldots,V$, where $V$ is the number of vertices --, and $\varepsilon_{vi}$ is the incidence matrix ($\varepsilon_{vi} = +1$ if the line $i$ starts at the vertex $v$, $\varepsilon_{vi} = -1$ if the line $i$ ends at the vertex $v$ and $\varepsilon_{vi}=0$ if the line $i$ is not related to the vertex $v$). To abbreviate notation, $\prod_{i}$ will always refer to $\prod_{i=1}^I$, unless specified.
	
	We introduce $d$ ($d<D$) periodically compactified dimensions by employing an extension of the Matsubara formalism of imaginary time~\cite{Khanna:2009zz,Khanna:2014qqa}. The prescription is well established to also consider the introduction of a chemical potential or to employ quasiperiodic/anyonic boundary conditions. However, for this work, we stick with just the periodic boundary conditions. We denote by a capital letter the full $D$-dimensional momenta ($K_i$) and by lower case the momenta related to the $D-d$ non-compactified dimensions, that is
	\begin{subequations}
		\begin{align}
		K_i^2 &\rightarrow k_i^2 + \sum_{\alpha} (\omega_\alpha^{k_i})^2,\\
		\int \frac{d^{D}K_i}{(2\pi)^{D}} &\rightarrow \int \frac{d^{D-d}k_i}{(2\pi)^{D-d}} \frac{1}{\prod_\alpha L_\alpha} \sum_{ \substack{n_\alpha^{k_i}\in \mathbb{Z} \\ \forall \alpha}}, \\
		\omega_\alpha^{k_i} &=  \frac{2\pi}{L_\alpha} n_\alpha^{k_i}.
		\end{align}
	\end{subequations}
	
	\noindent Where the index $\alpha$ runs over the $d$ compactified dimensions, $L_\alpha$ is the characteristic length for each of them and $\omega_\alpha^{k_i}$ is the $\alpha$-th Matsubara frequency associated with the momenta $k_i$. Notice that $\prod_\alpha$ must be undestood as $\prod_{\alpha=1}^d$. 
	
	Therefore, the amplitude for the graph $G$ in compactified dimensions turns to be
	\begin{multline}
	\widetilde{\mathcal{I}}_G = C_G \prod_{i} \left[\int \frac{d^{D-d} k_i}{(2\pi)^{D-d}} \frac{1}{\prod_\alpha L_\alpha} \sum_{ \substack{n_\alpha^{k_i}\in \mathbb{Z} \\ \forall \alpha}} 
	\frac{1}{k_i^2+\sum_\alpha \left(\omega_\alpha^{k_i}\right)^2+m_i^2}\right]\times\\
	\prod_{v=1}^V \left[(2\pi)^{D} \delta^{D-d} \left(p_v-\sum_{i} \varepsilon_{vi} k_i\right) \prod_\alpha \delta\left(\omega_\alpha^{p_v}-\sum_i \varepsilon_{vi}\omega_\alpha^{k_i}\right)\right].
	\end{multline}
	
	To proceed we start dealing with the non-compactified dimensions. At this point the procediment is closely related to the standard one, but is exhibited for completeness and clarity. We introduce a integral representation for the propagator and the conservation deltas, introducing the Schwinger parameters $u_i$ related to each internal line,
	\begin{multline}
	\widetilde{\mathcal{I}}_G = C_G \prod_{i} \left[\int \frac{d^{D-d} k_i}{(2\pi)^{D-d}} \frac{1}{\prod_\alpha L_\alpha} \sum_{ \substack{n_\alpha^{k_i}\in \mathbb{Z} \\ \forall \alpha}} 
	\int_0^\infty du_i e^{-u_i\left[k_i^2+\sum_\alpha \left(\omega_\alpha^{k_i}\right)^2+m_i^2\right]}
	\right]\times\\
	\prod_{v=1}^V \left[ \int d^{D-d}y_v e^{-iy_v\cdot\left(p_v-\sum_{i} \varepsilon_{vi} k_i\right)} \prod_\alpha 
	\left[
	\int d z_v^{(\alpha)}
	e^{-i z_v^{(\alpha)}\left(\omega_\alpha^{p_v}-\sum_i \varepsilon_{vi}\omega_\alpha^{k_i}\right) }
	\right]\right],
	\end{multline}
	\noindent we can interchange the sign of integrals and sums to make evident the integral over the internal momenta $k_i$ and also the sum over the internal frequencies $\omega_\alpha^{k_i}$, that is
	
	%\begin{multline}
	%\widetilde{\mathcal{I}}_G = C_G 
	%\left[\prod_i \int_0^\infty du_i\right]
	%\left[\prod_{v=1}^V \int d^{D-d}y_v\right]
	%\left[\prod_{v=1}^V \prod_\alpha \int d z_v^{(\alpha)}\right]
	%\prod_{i}\left[\int \frac{d^{D-d} k_i}{(2\pi)^{D-d}}\frac{1}{\prod_\alpha L_\alpha} \sum_{ \substack{n_\alpha^{k_i}\in \mathbb{Z} \\ \forall \alpha}} \right]
	%\times\\
	% e^{-\sum_{i }u_i\left[k_i^2+\sum_\alpha \left(\omega_\alpha^{k_i}\right)^2+m_i^2\right]}
	%e^{-i\sum_{v=1}^V \sum_\alpha z_v^{(\alpha)}\left(\omega_\alpha^{p_v}-\sum_i \varepsilon_{vi}\omega_\alpha^{k_i}\right) }
	%e^{-i\sum_{v=1}^Vy_v\cdot\left(p_v-\sum_{i} \varepsilon_{vi} k_i\right)}
	%\end{multline}
	
	\begin{multline}
	\widetilde{\mathcal{I}}_G = C_G 
	\left[\prod_i \int_0^\infty du_i\right]
	\left[\prod_{v=1}^V \int d^{D-d}y_v\right]
	\left[\prod_{v=1}^V \prod_\alpha \int d z_v^{(\alpha)}\right]
	e^{-\sum_{i }u_i m_i^2}
	e^{-i\sum_{v=1}^V \sum_\alpha z_v^{(\alpha)}\omega_\alpha^{p_v} }
	e^{-i\sum_{v=1}^Vy_v\cdot p_v}\times\\
	\prod_{i}\left[\int \frac{d^{D-d} k_i}{(2\pi)^{D-d}}
	e^{-u_i k_i^2}
	e^{i\sum_{v=1}^Vy_v\cdot \varepsilon_{vi} k_i}
	\right]
	\prod_{i}\left[\frac{1}{\prod_\alpha L_\alpha} \sum_{ \substack{n_\alpha^{k_i}\in \mathbb{Z} \\ \forall \alpha}} 
	e^{-u_i\sum_\alpha \left(\omega_\alpha^{k_i}\right)^2}
	e^{i\sum_{v=1}^V \sum_\alpha z_v^{(\alpha)}\varepsilon_{vi}\omega_\alpha^{k_i} }\right].
	\end{multline}
	
	Notice that the integral over $k_i$ is a gaussian and we can complete the squares and compute the integral to get
	\begin{equation}
	\int \frac{d^{D-d} k_i}{(2\pi)^{D-d}}
	e^{-u_i k_i^2}
	e^{i\sum_{v=1}^Vy_v\cdot \varepsilon_{vi} k_i}
	=
	\frac{e^{-\frac{\left(\sum_{v=1}^V y_v\varepsilon_{vi}\right)^2}{4u_i}}}{\left(4\pi u_i\right)^{\frac{D-d}{2}}},
	\end{equation}
	\noindent which substituted back into $\widetilde{\mathcal{I}}_G$ produce
	\begin{multline}
	\widetilde{\mathcal{I}}_G = C_G 
	\left[\prod_i \int_0^\infty du_i\right]
	\left[\prod_{v=1}^V \int d^{D-d}y_v\right]
	\left[\prod_{v=1}^V \prod_\alpha \int d z_v^{(\alpha)}\right]
	e^{-\sum_{i }u_i m_i^2}
	e^{-i\sum_{v=1}^V \sum_\alpha z_v^{(\alpha)}\omega_\alpha^{p_v} }
	e^{-i\sum_{v=1}^Vy_v\cdot p_v}\times\\
	\prod_{i}\left[\frac{e^{-\frac{\left(\sum_{v=1}^V y_v\varepsilon_{vi}\right)^2}{4u_i}}}{\left(4\pi u_i\right)^{\frac{D-d}{2}}}
	\right]
	\prod_{i}\left[\frac{1}{\prod_\alpha L_\alpha} \sum_{ \substack{n_\alpha^{k_i}\in \mathbb{Z} \\ \forall \alpha}} 
	e^{-u_i\sum_\alpha \left(\omega_\alpha^{k_i}\right)^2 + i\sum_{v=1}^V \sum_\alpha z_v^{(\alpha)}\varepsilon_{vi}\omega_\alpha^{k_i}}\right].
	\end{multline}
	
	Now we make evident the global conservation required by the delta function. To do so we employ the change of variables
	\begin{subequations}
		\begin{align}
		y_v = \bar y_v + y_V, &\qquad \forall v\neq V;\\
		y_V = \bar y_V; &\\
		z_v^{(\alpha)} = \bar z_v^{(\alpha)} + z_V^{(\alpha)}, &\qquad \forall v\neq V;\\
		z_V^{(\alpha)} = \bar z_V^{(\alpha)}. &
		\end{align}
		\noindent With this the sum over all vertices produce
		\begin{equation}
		\sum_{v=1}^V y_v \varepsilon_{vi} = \sum_{v=1}^{V-1} \bar{y}_v \varepsilon_{vi}
		+ \bar{y}_V \sum_{v=1}^V \varepsilon_{vi} = \sum_{v=1}^{V-1} \bar{y}_v \varepsilon_{vi},
		\end{equation}
	\end{subequations}
	
	\noindent because $\sum_{v=1}^V \varepsilon_{vi} =0$ as a property of the incidence matrix (each internal line starts from one vertex $+1$ and ends at one vertex $-1$, therefore the sum over all contributions is $0$). Applying this change of variables,
	
	\begin{multline}
	\widetilde{\mathcal{I}}_G = C_G 
	\left[\prod_i \int_0^\infty du_i\right]
	\left[\prod_{v=1}^V \int d^{D-d}\bar{y}_v\right]
	\left[\prod_{v=1}^V \prod_\alpha \int d \bar{z}_v^{(\alpha)}\right]
	e^{-\sum_{i }u_i m_i^2}
	e^{-i\sum_{v=1}^{V-1} \sum_\alpha \bar{z}_v^{(\alpha)}\omega_\alpha^{p_v} }
	e^{-i\sum_{v=1}^{V-1}\bar{y}_v\cdot p_v}\\
	e^{-i\sum_\alpha \bar{z}_V^{(\alpha)}\sum_{v=1}^{V} \omega_\alpha^{p_v} }
	e^{-i\bar{y}_V\cdot\sum_{v=1}^V p_v}
	\prod_{i}\left[\frac{e^{-\frac{\left(\sum_{v=1}^{V-1} \bar{y}_v\varepsilon_{vi}\right)^2}{4u_i}}}{\left(4\pi u_i\right)^{\frac{D-d}{2}}}
	\right]
	\prod_{i}\left[\frac{1}{\prod_\alpha L_\alpha} \sum_{ \substack{n_\alpha^{k_i}\in \mathbb{Z} \\ \forall \alpha}} 
	e^{-u_i\sum_\alpha \left(\omega_\alpha^{k_i}\right)^2 + i\sum_{v=1}^{V-1} \sum_\alpha \bar z_v^{(\alpha)}\varepsilon_{vi}\omega_\alpha^{k_i}}\right],
	\end{multline}
	
	\noindent one can extract the global conservation (for the $v=V$ component),
	\begin{equation}
	\int d^{D-d} \bar{y}_V \left[\prod_\alpha \int d\bar{z}_V^{(\alpha)} \right] e^{-i\sum_\alpha \bar{z}_V^{(\alpha)}\sum_{v=1}^{V} \omega_\alpha^{p_v} }
	e^{-i\bar{y}_V\cdot\sum_{v=1}^V p_v}
	=
	\delta^{D-d}\left(\sum_{v=1}^V p_v\right)
	\prod_\alpha \delta\left( \sum_{v=1}^{V} \omega_\alpha^{p_v} \right).
	\end{equation}
	
	Due to this decomposition, it is usual to define a new amplitude without the overall conservation,
	\begin{equation}
	\widetilde{\mathcal{I}}_G = 
	\delta^{D-d}\left(\sum_{v=1}^V p_v\right)
	\prod_\alpha \delta\left( \sum_{v=1}^{V} \omega_\alpha^{p_v} \right) \mathcal{I}_G,
	\end{equation}
	\noindent such that,
	\begin{multline}
	\mathcal{I}_G = C_G 
	\left[\prod_i \int_0^\infty du_i\right]
	\left[\prod_{v=1}^{V-1} \int d^{D-d}\bar{y}_v\right]
	\left[\prod_{v=1}^{V-1} \prod_\alpha \int d \bar{z}_v^{(\alpha)}\right]
	e^{-\sum_{i }u_i m_i^2}
	e^{-i\sum_{v=1}^{V-1} \sum_\alpha \bar{z}_v^{(\alpha)}\omega_\alpha^{p_v} }
	e^{-i\sum_{v=1}^{V-1}\bar{y}_v\cdot p_v} \times\\
	\prod_{i}\left[\frac{e^{-\frac{\left(\sum_{v=1}^{V-1} \bar{y}_v\varepsilon_{vi}\right)^2}{4u_i}}}{\left(4\pi u_i\right)^{\frac{D-d}{2}}}
	\right]
	\prod_{i}\left[\frac{1}{\prod_\alpha L_\alpha} \sum_{ \substack{n_\alpha^{k_i}\in \mathbb{Z} \\ \forall \alpha}} 
	e^{-u_i\sum_\alpha \left(\omega_\alpha^{k_i}\right)^2 + i\sum_{v=1}^{V-1} \sum_\alpha \bar z_v^{(\alpha)}\varepsilon_{vi}\omega_\alpha^{k_i}}\right].
	\end{multline}
	
	%\begin{multline}
	%\mathcal{I}_G = C_G 
	%\left[\prod_i \int_0^\infty du_i\right]
	%e^{-\sum_{i }u_i m_i^2}
	%\prod_{i}\left[\frac{1}{\left(4\pi u_i\right)^{\frac{D-d}{2}}}
	%\right]
	%\prod_{i}\left[\frac{1}{\prod_\alpha L_\alpha} \sum_{ \substack{n_\alpha^{k_i}\in \mathbb{Z} \\ \forall \alpha}} \right]
	%\times\\
	%\prod_\alpha \left\{
	% \left[ \prod_{v=1}^{V-1} 
	%\int d \bar{z}_v^{(\alpha)}
	%\right]
	%e^{-i \sum_{v=1}^{V-1} \bar{z}_v^{(\alpha)}\omega_\alpha^{p_v} }
	%e^{-\sum_i u_i \left(\omega_\alpha^{k_i}\right)^2 + i\sum_{v=1}^{V-1} \sum_i \bar z_v^{(\alpha)}\varepsilon_{vi}\omega_\alpha^{k_i}}
	%\right\}
	% \\
	%\left[\prod_{v=1}^{V-1} \int d^{D-d}\bar{y}_v\right] 
	%e^{-i\sum_{v=1}^{V-1}\bar{y}_v\cdot p_v} e^{-\sum_i\frac{\left(\sum_{v=1}^{V-1} \bar{y}_v\varepsilon_{vi}\right)^2}{4u_i}}
	%\end{multline}
	
	At this point we compute the gaussian integral over the parameter $\bar y_v$,
	\begin{equation}
	\left[\prod_{v=1}^{V-1} \int d^{D-d}\bar{y}_v\right] 
	e^{-i\sum_{v=1}^{V-1}\bar{y}_v\cdot p_v} e^{-\sum_i\frac{\left(\sum_{v=1}^{V-1} \bar{y}_v\varepsilon_{vi}\right)^2}{4u_i}}
	=
	\frac{(4\pi)^{\frac{(D-d)}{2}(V-1) }}{\left[\det d_G(u)\right]^{\frac{D-d}{2}}} e^{- \sum_{v_1,v_2=1}^{V-1}p_{v_1}p_{v_2}\left[d^{-1}_G(u)\right]_{v_1,v_2}}.
	\end{equation}
	Here, it was defined the symmetric $(V-1)\times(V-1)$ matrix $d_G(u)$ as
	\begin{equation}
	\left[d_G(u)\right]_{v_1,v_2} = \sum_i \frac{\varepsilon_{v_1i}\varepsilon_{v_2i}}{u_i}.
	\label{Eq:def_dG}
	\end{equation}
	
	This produces, after a bit of organization,
	\begin{multline}
	\mathcal{I}_G = C_G 
	\left[\prod_i \int_0^\infty du_i\right]
	\frac{e^{-\sum_{i }u_i m_i^2}}{(4\pi)^{\frac{(D-d)}{2}L}}
	\frac{e^{-\frac{V(p)}{U}}}{U^{\frac{D-d}{2}}} 
	\prod_{i}\left[\frac{1}{\prod_\alpha L_\alpha} \sum_{ \substack{n_\alpha^{k_i}\in \mathbb{Z} \\ \forall \alpha}} \right]\\
	\prod_\alpha \left\{
	\left[ \prod_{v=1}^{V-1} 
	\int d \bar{z}_v^{(\alpha)}
	\right]
	e^{-i \sum_{v=1}^{V-1} \bar{z}_v^{(\alpha)}\omega_\alpha^{p_v} }
	e^{-\sum_i u_i \left(\omega_\alpha^{k_i}\right)^2 + i\sum_{v=1}^{V-1} \sum_i \bar z_v^{(\alpha)}\varepsilon_{vi}\omega_\alpha^{k_i}}\right\},
	\label{Eq:IG_pretratamento}
	\end{multline}
	\noindent or, introducing back the delta functions for the compactified dimensions,
	\begin{multline}
	\mathcal{I}_G = C_G 
	\left[\prod_i \int_0^\infty du_i\right]
	\frac{e^{-\sum_{i }u_i m_i^2}}{(4\pi)^{\frac{(D-d)}{2}L}}
	\frac{e^{-\frac{V(p)}{U}}}{U^{\frac{D-d}{2}}} 
	\prod_{i}\left[\frac{1}{\prod_\alpha L_\alpha} \sum_{ \substack{n_\alpha^{k_i}\in \mathbb{Z} \\ \forall \alpha}} e^{-\sum_\alpha u_i \left(\omega_\alpha^{k_i}\right)^2}
	\right]
	\prod_{v=1}^{V-1} \prod_\alpha (2\pi)\delta\left(
	\omega_\alpha^{p_v} - \sum_i \varepsilon_{vi}\omega_\alpha^{k_i}
	\right).
	\end{multline}
	\noindent Where $L = I-V+1$ is the number of loops in the diagram (do not confuse with $L_\alpha$, the length of each compactified dimension), and the polynomials $U$ and $V$ are the Symanzik polynomials~\cite{Itzykson:1980rh,Rivasseau:1991ub}, defined here as
	\begin{subequations}
		\label{Eq:Symanzik}
		\begin{align}
		\frac{V(q;u)}{U(u)} &= \sum_{v_1,v_2=1}^{V-1}q_{v_1}q_{v_2}\left[d^{-1}_G(u)\right]_{v_1,v_2}, \label{Eq:V}\\
		U(u) &= \left(\prod_i u_i\right)\det d_G(u). \label{Eq:U}
		\end{align}
	\end{subequations}
	\noindent These polynomials have the remarkable propertie that they can be obtained directly from the graph $G$ by inspecting the 1-trees and 2-trees that can be formed by removing some internal lines. That is,
	\begin{subequations}
		\label{Eq:SymanzikTree}
		\begin{align}
		U &= \sum_T \prod_{i \notin T} u_i\\
		V &= \sum_K Q_K^2 \prod_{i\notin K} u_i
		\end{align}
	\end{subequations}
	\noindent meaning that the first polynomial $U$ is the sum over all 1-trees $T$ (connected graphs without loops) and we consider all lines $i$ that are not in the 1-tree, the removed lines. Also, the second polynomial $V$ is the sum over all 2-trees $K$ (two separated trees) where we take all parameters $u_i$ that do not belong to the 2-tree. $Q_K$ is the overall momenta that enters the 2-tree.
	
	This ends the application of the usual procedure to the non-compactified dimensions. From now on we deal with the remaining $d$ dimensions. To proceed we employ the delta functions (in fact Kronecker deltas) to reduce the number of summations just to the loop summations and then treat the expression. %There is also another path, starting from Eq~\eqref{Eq:IG_pretratamento}, where all sums are maintained. This path is shown in the App.~\ref{App:Path2}, it has the advantage to show explicitly the appearance of the Symanzik polynomials also in the compactified sector. However, it has the drawback that the sum over all internal lines is kept, which turns a bit difficult and odd the direct use of this representation for computation. We point out that the path followed in the appendix, although not the same, resembles what has been done by Ref.~\cite{Benhamou:1990cd,Benhamou:1992} in finite temperature field theory.
	
	From this point forward $\prod_v$ refers to $\prod_{v=1}^{V-1}$ unless otherwise specified. As the sum is over the modes, one can extract a factor from the delta functions
	
	\begin{equation}
	\mathcal{I}_G = C_G 
	\left[\prod_i \int_0^\infty du_i\right]
	\frac{e^{-\sum_{i }u_i m_i^2}}{(4\pi)^{\frac{(D-d)}{2}L}}
	\frac{e^{-\frac{V(p)}{U}}}{U^{\frac{D-d}{2}}}
	\prod_\alpha\left\{
	\frac{1}{L_\alpha^I} \sum_{ \substack{n_\alpha^{k_i}\in \mathbb{Z}\\\forall i}} e^{- \sum_i u_i \left(\omega_\alpha^{k_i}\right)^2}
	\left(L_\alpha\right)^{V-1}\prod_v \delta\left(
	n_\alpha^{p_v} - \sum_i \varepsilon_{vi}n_\alpha^{k_i}
	\right)
	\right\}.
	\label{Eq:IG_withdeltas}
	\end{equation}
	
	\noindent Notice that for each $\alpha$-th Matsubara mode we have $I$ summations (related to the internal lines) and $V-1$ relations between the frequencies (given by the Kronecker deltas), meaning an overall $L=I+1-V$ free frequencies to be summed up. This indicates that the notation could be changed to something like
	\begin{equation}
	\frac{1}{L_\alpha^L}\prod_v \sum_{ \substack{n_\alpha^{k_i}\in \mathbb{Z}\\\forall i}} e^{- \sum_i u_i \left(\omega_\alpha^{k_i}\right)^2}
	\prod_v \delta\left(
	n_\alpha^{p_v} - \sum_i \varepsilon_{vi}n_\alpha^{k_i}\right)
	=
	\frac{1}{L_\alpha^L}e^{-Z[\omega_\alpha^{p_v}]}
	\sum_{ \substack{n_\alpha^{(\ell)}\in \mathbb{Z}\\\forall \ell \in [1,L]}} e^{- Y[\omega_\alpha^{k_\ell}]},
	\end{equation}
	\noindent where the index $\ell$ runs over the $L$ independent loops. $Y[\omega_\alpha^{k_\ell}]$ is a bilinear form on the $L$ frequencies related to the loops, $Y$ is a $L\times L$ matrix whose coefficients depend on $u_i$, the external modes $n_\alpha^{p_v}$ and the incidence matrix. Also, $Z[\omega_\alpha^{p_v}]$ is another bilinear form that depends on the external modes. Let us now proceed to obtain it using the delta functions, we need to rewrite
	\begin{equation}
	\Delta_\alpha = - \sum_i u_i \left(\omega_\alpha^{k_i}\right)^2
	\label{Eq:ExpDelta}
	\end{equation}
	\noindent using the $V-1$ constraints
	\begin{equation}
	\omega_\alpha^{p_v} - \sum_i \varepsilon_{vi}\omega_\alpha^{k_i} = 0.
	\end{equation}
	\noindent We choose a prescription where the \textit{last lines} will be eliminated, and we use the constraints to define a linear system,
	\begin{equation}
	\omega_\alpha^{p_v} - \sum_{\ell} \varepsilon_{v\ell}\omega_\alpha^{k_\ell} = \sum_{v_0=1}^{V-1} \varepsilon_{v,L+v_0}\omega_\alpha^{k_{L+v_0}},
	\end{equation}
	\noindent that determine the frequencies to be eliminated ($\omega_\alpha^{k_{L+v_0}}$) as a function of the $L$ frequencies to be kept ($\omega_\alpha^{k_\ell}$). In matricial notation, this means trivally that
	\begin{subequations}
		\begin{equation}
		Q = \bar{\epsilon} \overline{W} \Rightarrow \overline{W} = \bar{\epsilon}^{-1} Q,
		\label{Eq:Wsolution}
		\end{equation}
		\noindent with
		\begin{align}
		(Q)_v = \omega_\alpha^{p_v} - \sum_{\ell} \varepsilon_{v\ell}\omega_\alpha^{k_\ell} \label{Eq:Q},\\
		(\bar{\epsilon})_{v,v_0} = \varepsilon_{v,L+v_0},\\
		(\overline{W})_{v_0} = \omega_\alpha^{k_{L+v_0}}.
		\end{align}
	\end{subequations}
	
	Of course, for this to work, we need the $(V-1)\times(V-1)$ matrix $\bar \epsilon$ to be invertible. This is not guaranteed for all choices of labeling of lines and vertices when building the incidence matrix $\varepsilon_{v,i}$. But we can always reorganize the labels to guarantee a choice where $\bar \epsilon$ is invertible. There are a few remarks here. First, here we always choose to eliminate the last internal lines, this choice does not affect the results, is just a matter of aesthetics (\textit{we want to keep the same labels for the number of loops, so the $L$ frequencies related to the $L$ loops are exactly the first $L$ internal lines}). Second, different choices of $\bar \epsilon$ will not affect the appearance of the Symanzik polynomials $U$ and $V$, as one should expect for consistency. However, the matrix $Y$ will depend on this choice. Anyway, one can show that through a suitable rearrangement of the modes ($n_\alpha^{k_\ell}$) one can relate a matrix $Y$ obtained by some $\bar \epsilon$ to another matrix $Y$ obtained by another choice of $\bar \epsilon$. Therefore, we do not lose generality by specifying some prescriptions to obtain $\bar \epsilon$.
	
	If we substitute back Eq.~\eqref{Eq:Wsolution} into Eq.~\eqref{Eq:ExpDelta},
	\begin{equation*}
	\Delta_\alpha = - \sum_\ell u_\ell \left(\omega_\alpha^{k_\ell}\right)^2
	- \sum_{v_0=1}^{V-1} u_{L+j} \left(\omega_\alpha^{k_{L+j}}\right)^2
	=
	- \sum_\ell u_\ell \left(\omega_\alpha^{k_\ell}\right)^2
	- \sum_{v_0,v_1,v_2=1}^{V-1} u_{L+j} \left[(\bar{\epsilon}^{-1})_{v_0,v_1}Q_{v_1}\right]
	\left[(\bar{\epsilon}^{-1})_{v_0,v_2}Q_{v_2}\right],
	\end{equation*}
	\noindent and then apply the expression for $Q_{v}$, Eq.~\eqref{Eq:Q}, we obtain after some algebraic manipulations that, in matricial notation,
	\begin{subequations}
		\label{Eq:MBF_def}
		\begin{equation}
		\Delta_\alpha = -(W_\alpha-M^{-1}B_\alpha)^t M (W_\alpha-M^{-1}B_\alpha) + (-F_\alpha + B_\alpha^t M^{-1} B_\alpha),
		\end{equation}
		\noindent where $W_\alpha,M,B_\alpha$ and $F_\alpha$ are:
		\begin{align}
		(W_\alpha)_\ell &= \omega_\alpha^{k_\ell}, \label{Eq:W_def}\\
		(M)_{\ell_1,\ell_2} &= u_{\ell_1} \delta_{\ell_1,\ell_2} + \sum_{v_0,v_1,v_2} u_{L+v_0}\left[(\bar\epsilon^{-1})_{v_0,v_1}\varepsilon_{v_1,\ell_1}\right]\left[(\bar\epsilon^{-1})_{v_0,v_2}\varepsilon_{v_2,\ell_2}\right]\label{Eq:M_def},\\
		(B_\alpha)_\ell &= \sum_{v_0,v_1,v_2} u_{L+v_0}\left[(\bar\epsilon^{-1})_{v_0,v_1}\varepsilon_{v_1,\ell}\right]\left[(\bar\epsilon^{-1})_{v_0,v_2}\omega_\alpha^{P_{v_2}}\right], \label{Eq:B_def} \\
		F_\alpha &= \sum_{v_0,v_1,v_2} u_{L+v_0}\left[(\bar\epsilon^{-1})_{v_0,v_1}\omega_\alpha^{P_{v_1}}\right]\left[(\bar\epsilon^{-1})_{v_0,v_2}\omega_\alpha^{P_{v_2}}\right].\label{Eq:F_def}
		\end{align}
		\noindent Here, $M$ is a $L\times L$ matrix. For convience, we can also define a \textit{tilde} notation to indicate that the factor $2\pi/L_\alpha$ is extracted,
		\begin{align}
		(\widetilde W_\alpha)_\ell &= n_\alpha^{k_\ell},\label{Eq:Wtilde_def}\\
		(\widetilde B_\alpha)_\ell &= \sum_{v_0,v_1,v_2} u_{L+v_0}\left[(\bar\epsilon^{-1})_{v_0,v_1}\varepsilon_{v_1,\ell}\right]\left[(\bar\epsilon^{-1})_{v_0,v_2}n_\alpha^{P_{v_2}}\right]. \label{Eq:Btilde_def}
		\end{align}
	\end{subequations}
	
	At this point we can get back to the original expression, Eq.~\eqref{Eq:IG_withdeltas}, and rewrite the amplitude related to some graph $G$ as, 
	\begin{equation}
	\mathcal{I}_G = C_G 
	\left[\prod_i \int_0^\infty du_i\right]
	\frac{e^{-\sum_{i }u_i m_i^2}}{(4\pi)^{\frac{(D-d)}{2}L}}
	\frac{e^{-\frac{V(p)}{U}}}{U^{\frac{D-d}{2}}} \prod_\alpha\left\{\frac{1}{L_\alpha^L}
	e^{-F_\alpha+B_\alpha^tM^{-1}B_\alpha}
	\sum_{ \substack{n_\alpha^{(\ell)}\in \mathbb{Z}\\\forall \ell \in [1,L]}} e^{-\frac{4\pi^2}{L_\alpha^2}(\widetilde W_\alpha-M^{-1}\widetilde B_\alpha)^t M (\widetilde W_\alpha-M^{-1}\widetilde B_\alpha)}
	\right\}.
	\label{Eq:IG_compactified}
	\end{equation}
	
	With the previous procedure (integrate over the momenta and then integrate over the parameters related to the delta function) we found the standard definition of the Symanzik polynomials, see Eq.~\eqref{Eq:Symanzik}, in terms of the incidence matrix. Using the alternative procedure (apply first the delta function and reduce the number of integrals) the Symanzik polynomials can be expressed as~\footnote{One can check it repeating the procedure for the non-compact dimensions. This is discussed in Ref.~\cite{Todorov:1971book}.}
	
	\begin{subequations}
		\label{Eq:SymanzikTodorov}
		\begin{align}
		U(u)&=\det M,\\
		\frac{V(\omega_\alpha^{p},u)}{U(u)} &= F_\alpha-B_\alpha^tM^{-1}B_\alpha.
		\end{align}
	\end{subequations}
	
	\noindent With $M$, $B$ and $F$ defined as in Eq.~\eqref{Eq:MBF_def}. Therefore, one can simplify the representation to 
	
	\begin{equation}
	\mathcal{I}_G = C_G 
	\left[\prod_i \int_0^\infty du_i\right]
	\frac{e^{-\sum_{i }u_i m_i^2}}{(4\pi)^{\frac{(D-d)}{2}L}}
	\frac{e^{-\frac{V(P)}{U}}}{U^{\frac{D-d}{2}}} \prod_\alpha\left\{\frac{1}{L_\alpha^L}
	\sum_{ \substack{n_\alpha^{(\ell)}\in \mathbb{Z}\\\forall \ell \in [1,L]}} e^{-\frac{4\pi^2}{L_\alpha^2}(\widetilde W_\alpha-M^{-1}\widetilde B_\alpha)^t M (\widetilde W_\alpha-M^{-1}\widetilde B_\alpha)}
	\right\},
	\label{Eq:IG_compactifiedsimplify}
	\end{equation}
	\noindent remember that $P_v=(p_v,\omega_\alpha^{p_v})$ is the full external momenta, meaning that $V(P) = V(p) + \sum_\alpha V(\omega_\alpha^{p},u)$.
	
	From this point forward we will consider two different scenarios: the regime near dimensional reduction ($L_\alpha \rightarrow 0$) and the regime near bulk ($L_\alpha \rightarrow \infty$). Both of them are equivalent analytically, in the sense that it is possible to transport from one to another, but each of them is more suitable for a different length scale. At first, we show the behavior when the compactification lengths are small (near a dimensional reduction), in this scenario the expression is exactly the one from Eq.~\eqref{Eq:IG_compactified} but requires some treatment, this is done in Sec.~\ref{Sec:NearRedDim}. Then, in Sec.~\ref{Sec:NearBulk}, we show the behavior when the compactification lengths are large (near the bulk), for this we employ a Jacobi theta identification that modifies the summation in Eq.~\eqref{Eq:IG_compactified}. The difference between both procedures is that, although both are indeed valid for all compactified sizes, each one converges faster on the specified limit. That said, although it can be done in principle it is not indicated to employ the representation from Sec.~\ref{Sec:NearRedDim} to study the bulk limit neither the representation from Sec.~\ref{Sec:NearBulk} to study the dimensional reduction limit.
	
	\section{Near dimensional reduction representation}\label{Sec:NearRedDim}
	
	As far as the author knows, although the topic of dimensional reduction is well-known in the subject of finite temperature field theory (a dimensional reduction occurs in the very high-temperature limit) the only attempt to understand the behavior of a Feynman amplitude with many periodically compactified spaces near a dimensional reduction comes from \cite{Cavalcanti:2018pgi,Cavalcanti:2019mli}. The present context, where we produce a parametric representation for a scalar field theory in a periodically compactified space, allows a clear and easier evaluation of the dimensional reduction.
	
	At first, let us make clear that we consider the dimensional reduction in the sense of Fisher~\cite{Fisher:1973jvst}. That is, one does not say that all length parameters are zero ($L_\alpha=0$), but rather that the length is small enough so that its contribution can be mostly ignored. So we can take the limit where $L_\alpha\rightarrow 0$ and keeps track only of the dominant contribution.
	
	Taking the amplitude $I_G$ as in  Eq.~\eqref{Eq:IG_compactified}, we see that the dominant contribution from the summation comes from a combination of modes $n_\alpha^{(\ell)}$ that minimize the exponential suppression. All other internal modes are responsible for subdominant contributions in this limit. The dominant contribution is 
	
	\begin{equation}
	\mathcal{I}_G \sim C_G 
	\left[\prod_i \int_0^\infty du_i\right]
	\frac{e^{-\sum_{i }u_i m_i^2}}{(4\pi)^{\frac{(D-d)}{2}L}}
	\frac{e^{-\frac{V(p)}{U}}}{U^{\frac{D-d}{2}}}
	\prod_\alpha\left[ \frac{e^{-\frac{4\pi^2}{L_\alpha^2} \min G_\alpha}}{L_\alpha^L}\right].
	\end{equation}
	\noindent Where $G_\alpha$ depends on the parameters $u_i$ and the external modes $n_\alpha^{p_v}$ as
	\begin{equation}
	G_\alpha =
	\widetilde{F}_\alpha - \widetilde B_\alpha^tM^{-1} \widetilde B_\alpha
	+ (\widetilde W_\alpha-M^{-1}\widetilde B_\alpha)^t M (\widetilde W_\alpha-M^{-1}\widetilde B_\alpha)
	\end{equation}
	\noindent and the minimization is with respect to the internal modes. That is, one chooses a set $\widetilde W_\alpha$ such that $G_\alpha$ has its minimum value. If $M^{-1}\widetilde B_\alpha=0$ this is trivially $\widetilde W_\alpha=0$. The component $L_\alpha^L$ can be easily absorbed by the definition of the coupling constant (that is inside the constant $C_G$), but $G_\alpha$ is not easily absorbed. This component introduces a new contribution to the Symanzik polynomials. That is, in the limit of dimensional reduction one do not simply get the same amplitude in reduced dimensions, but there is some surviving information from the original dimensions.
	
	There is another perspective, where one considers a completely static mode approach. In this perspective all modes, internal and external, are null, and this eliminates the contribution $G_\alpha$ producing simply
	\begin{equation}
	\mathcal{I}_G \Big|_{\{n_\alpha^{k_\ell},n_\alpha^{p_v}\}=0}= \frac{C_G}{\prod_\alpha L_\alpha^L} 
	\left[\prod_i \int_0^\infty du_i\right]
	\frac{e^{-\sum_{i }u_i m_i^2}}{(4\pi)^{\frac{(D-d)}{2}L}}
	\frac{e^{-\frac{V(p)}{U}}}{U^{\frac{D-d}{2}}},
	\end{equation}
	\noindent notice we can absorb the remaining factor related with the size in the constant $C_G$ associated with the vertices. This is exactly the usual parametric representation of Feynman diagram but only considering the $D-d$ non-compactified and large dimensions.
	
	Sometimes it might look that both procedures are equivalent. The difference between both becomes more evident if we consider a scenario where the graph $G$ is a subdiagram to be evaluated in the ultraviolet regime, in this scenario the external momenta are large and there is no justification to use a zero-mode approximation for the external frequencies. This simple remark can be responsible to produce large differences in behavior.
	
	The remaining terms, subdominant contributions, to the amplitude in the small length regime are obtained in straightforward way. Just as the dominant contribution is the minimum with respect to the internal modes ($G_\alpha$), the next leading order is a small deviation around this minimum. To make notation clear we apply a shift in the internal modes, that is
	\begin{equation}
	n_\alpha^{(\ell)} \rightarrow n_\alpha^{(\ell)} + j_\alpha^\ell,
	\end{equation}
	\noindent where $j_\alpha^\ell\in \mathbb{Z}$, this is chosen such that the minimum now lie on $n_\alpha^{(\ell)} = 0$. Let us write this shift as
	\begin{equation}
	\widetilde{W}_\alpha \rightarrow \widetilde{W}_\alpha - \widetilde{Y}_\alpha = \widetilde{W}_\alpha + M^{-1}\widetilde B_\alpha - \widetilde{Z}_\alpha,
	\end{equation}
	\noindent so that $G_\alpha$ can be rewritten as
	\begin{equation}
	G_\alpha =
	\widetilde{F}_\alpha - \widetilde B_\alpha^tM^{-1} \widetilde B_\alpha
	+ (\widetilde W_\alpha-\widetilde Z_\alpha)^t M (\widetilde W_\alpha-\widetilde Z_\alpha).
	\end{equation}
	\noindent We must choose a vector $\widetilde Y_\alpha\in\mathbb{Z}^L$ such that the components of the resultant vector $\widetilde Z_\alpha$ all lies in the interval $(\widetilde Z_\alpha)_\ell \in \left[-\frac{1}{2},\frac{1}{2}\right]$. This guarantees that the minimum value of $G_\alpha$ occurs at $\widetilde W_\alpha=0$. The simple choice is that $\widetilde Y_\alpha$ is the nearest integer to the real value $M^{-1} \widetilde B_\alpha$, which we denote by $\round{M^{-1} \widetilde B_\alpha}$. Therefore, the vector $\widetilde Z_\alpha$ is just
	\begin{equation}
	\widetilde Z_\alpha = M^{-1} \widetilde B_\alpha - \round{M^{-1} \widetilde B_\alpha}.
	\end{equation}
	
	With this simple change of notation the order of dominance is easier to write. The dominant contribution, as already stated, occur at $\sum_\ell |n_\alpha^{(\ell)}| = 0$, meaning that $n_\alpha^{(\ell)} = 0,\,  \forall \ell$. The first correction occurs at $\sum_\ell |n_\alpha^{(\ell)}| = 1$ and so on. This produces the simple expression,
	
	\begin{multline}
	\mathcal{I}_G = C_G 
	\left[\prod_i \int_0^\infty du_i\right]
	\frac{e^{-\sum_{i }u_i m_i^2}}{(4\pi)^{\frac{(D-d)}{2}L}}
	\frac{e^{-\frac{V(P)}{U}}}{U^{\frac{D-d}{2}}} \prod_\alpha\left\{\frac{1}{L_\alpha^L}
	e^{-Z_\alpha^t M Z_\alpha}
	\right\}\\
	+C_G 
	\left[\prod_i \int_0^\infty du_i\right]
	\frac{e^{-\sum_{i }u_i m_i^2}}{(4\pi)^{\frac{(D-d)}{2}L}}
	\frac{e^{-\frac{V(P)}{U}}}{U^{\frac{D-d}{2}}} \prod_\alpha\left\{\frac{1}{L_\alpha^L}
	\sum^{'}_{ \substack{n_\alpha^{(\ell)}\in \mathbb{Z}\\\forall \ell \in [1,L]}} e^{-\frac{4\pi^2}{L_\alpha^2}(\widetilde W_\alpha-\widetilde Z_\alpha)^t M (\widetilde W_\alpha-\widetilde Z_\alpha)}
	\right\}.
	\end{multline}
	\noindent Where $\sum'$ denotes that the zero mode is already removed. 
	
	In the next section, we deal with the scenario near the bulk (large compactification length) and rewrite the parametrization using a new distribution of sectors that change the $u$ parameters to $t$ parameters. Near the bulk, this reparametrization has the advantage of letting us see the appearance of the modified Bessel function of the second kind, characteristic of periodically compactified problems. However, when discussing the scenario near dimensional reduction this new parametrization does not contribute a lot, so we let the explanation of it for the next section. For completeness, however, we show below how the amplitude looks like after this reparametrization.
	
	\begin{multline}
	\mathcal{I}_G = \frac{1}{(\prod_\alpha L_\alpha)^L} C_G 
	\left[\prod_{i=1}^{I-1} \int_0^1 dt_i t_i^{I-1-i}\right] 
	\frac{\Gamma\left(I-\frac{(D-d)L}{2}\right)}{(4\pi)^{\frac{(D-d)}{2}L}{\overline U}^{\frac{(D-d)}{2}}}
	\frac{1}{\left[\mathcal{M}^2(t_i,m_i) + \frac{\overline V(P;t)}{\overline U(t)}
		+ {\overline Z}_\alpha^t \overline M  \overline Z_\alpha
		\right]^{I-\frac{(D-d)L}{2}}} 
	\\
	+ \frac{1}{(\prod_\alpha L_\alpha)^L} C_G 
	\left[\prod_{i=1}^{I-1} \int_0^1 dt_i t_i^{I-1-i}\right]  \frac{\Gamma\left(I-\frac{(D-d)L}{2}\right)}{(4\pi)^{\frac{(D-d)}{2}L}{\overline U}^{\frac{(D-d)}{2}}}
	\times\\
	\sum^{'}_{ \substack{n_\alpha^{(\ell)}\in \mathbb{Z}\\\forall \ell, \forall \alpha}}
	\frac{1}{\left[\mathcal{M}^2(t_i,m_i) + \frac{\overline V(P;t)}{\overline U(t)}
		+ (\overline W_\alpha - {\overline Z}_\alpha)^t \overline M  (\overline W_\alpha - \overline Z_\alpha)
		\right]^{I-\frac{(D-d)L}{2}}} 
	\label{Eq:Amplitude_NearRedDim}
	\end{multline}
	
	The remaining sum is a zeta-like summation, in the scenario of very small compactification lengths, it can be truncated at low terms (as discussed before, each new contribution is subdominant).
	
	\section{Near bulk representation} \label{Sec:NearBulk}
	
	Starting from Eq.~\eqref{Eq:IG_compactified} we apply the Jacobi theta identity,
	\begin{equation}
	\sum_{\mathbf{a} \in \mathbb{Z}^n} e^{-\pi t Y[\mathbf{a}+\mathbf{g}]} e^{2\pi i \mathbf{h} \cdot \mathbf{a}}.
	= e^{-2\pi i \mathbf{h} \cdot \mathbf{g}}  
	t^{-\frac{n}{2}} |Y|^{-\frac{1}{2}}
	\sum_{\mathbf{a}\in \mathbb{Z}^n} 
	e^{-\frac{\pi}{t} Y^{-1}[\mathbf{a-h}]}
	e^{2\pi i \mathbf{a} \cdot \mathbf{g}},
	\end{equation}
	\noindent with $t=4\pi/L_\alpha^2$, $n=L$, the matrix $Y=M$, $(\textbf{g})_n = -(M^{-1}\widetilde{B}_\alpha)_\ell$, $(\textbf{a})_n=(\widetilde W_\alpha)_\ell=n_\alpha^{k_\ell}$. So that we obtain
	
	\begin{equation}
	\mathcal{I}_G = C_G 
	\left[\prod_i \int_0^\infty du_i\right]
	\frac{e^{-\sum_{i }u_i m_i^2}}{(4\pi)^{\frac{(D-d)}{2}L}}
	\frac{e^{-\frac{V(P)}{U}}}{U^{\frac{D-d}{2}}} 
	\prod_\alpha
	\left\{
	\frac{
		1
	}{
		(4\pi)^{\frac{L}{2}}
		\left[\det M\right]^{\frac{1}{2}}
	} 
	\sum_{ \substack{n_\alpha^{(\ell)}\in \mathbb{Z}\\\forall \ell \in [1,L]}}
	e^{
		-\frac{L_\alpha^2}{4} \widetilde W_\alpha^t M^{-1} \widetilde W_\alpha
	}
	e^{
		2\pi i \widetilde W_\alpha^t M^{-1} \widetilde B_\alpha
	}
	\right\}.
	\end{equation}
	Recall the definition of the Symanzik polynomials $U$ and $V$ as in Eq.~\eqref{Eq:SymanzikTodorov}, the amplitude for a graph $G$ when $d$ dimensions are compactified can then be written as
	\begin{multline}
	\mathcal{I}_G = C_G 
	\left[\prod_i \int_0^\infty du_i\right]
	\frac{e^{-\sum_{i }u_i m_i^2}}{(4\pi)^{\frac{D}{2}L}}
	\frac{e^{-\frac{V(P;u)}{U(u)}}}{U(u)^{\frac{D}{2}}} 
	\times\\
	\sum_{ \substack{n_\alpha^{(\ell)}\in \mathbb{Z}\\\forall \ell, \forall \alpha}}
	e^{
		- \sum_\alpha \frac{L_\alpha^2}{4 U(u)} \sum_{\ell_1,\ell_2} n_\alpha^{k_{\ell_1}} n_\alpha^{k_{\ell_2}} A_{\ell_1,\ell_2}(u)
	}
	e^{
		\frac{2\pi i}{U(u)} \sum_{\alpha,\ell_1,\ell_2} n_\alpha^{k_{\ell_1}} A_{\ell_1,\ell_2}(u) (\widetilde{B}_\alpha(u))_{\ell_2}
	},
	\label{Eq:IG_nearbulk}
	\end{multline}
	\noindent or, in compact notation,
	\begin{equation}
	\mathcal{I}_G = C_G 
	\left[\prod_i \int_0^\infty du_i\right]
	\frac{e^{-\sum_{i }u_i m_i^2}}{(4\pi)^{\frac{D}{2}L}}
	\frac{e^{-\frac{V(P;u)}{U(u)}}}{U(u)^{\frac{D}{2}}} 
	\sum_{ \substack{n_\alpha^{(\ell)}\in \mathbb{Z}\\\forall \ell, \forall \alpha}}
	e^{
		- \sum_\alpha \frac{L_\alpha^2}{4 U(u)} \textbf{n}_\alpha^{t} A(u) \textbf{n}_\alpha
	}
	e^{
		\frac{2\pi i}{U(u)} \sum_{\alpha} \textbf{n}_\alpha^t A(u) \widetilde{B}_\alpha(u)
	}.
	\end{equation}
	\noindent here $\textbf{n}_\alpha$ is vector with components $(\textbf{n}_\alpha)_\ell = n_\alpha^{k_\ell}$ and the matrix $A$ is $A = M^{-1} \det M$. 
	
	Based on all we have done so far, the prescription one must follow is:
	\begin{enumerate}
		\item From the graph extract the 1-trees and produce $U(u)$ as usual, see Eq.~\eqref{Eq:SymanzikTree};
		\item From the graph extract the 2-trees and produce $V(u)$ as usual, see Eq.~\eqref{Eq:SymanzikTree};
		\item From the incidence matrix $\varepsilon_{v,i}$ extract a $(V-1)\times (V-1)$ matrix $\bar\epsilon$ that is invertible. By convention, take the last internal lines;
		\item Using $\bar\epsilon$, $\varepsilon_{v,i}$ and $\omega_\alpha^{p_v}$ compute the $M$, Eq.~\eqref{Eq:M_def}, and $\widetilde{B}$, Eq.~\eqref{Eq:Btilde_def}. The matrix $A$ is defined as $A = M^{-1} \det M$.
	\end{enumerate}
	
	Unfortunately, up to this point, we do not have a prescription that allows us to recover the matrix $A$ or the vector $\widetilde{B}$ directly from the topology of diagram, like $U$ and $V$ that are extracted by drawing the 1-trees and 2-trees related to the graph, see Eq.~\eqref{Eq:SymanzikTree}.
	
	The Symanzyk polynomials $U$ and $V$ are solely related to the graph, however, we also need $A$ and $B$, which are determined from our prescription of $\bar \epsilon$. One might inquire whether they are different depending on the choice of prescription and indeed they are. Unlike $U$ and $V$, which do not feel our choice of $\bar \epsilon$, the factors $A$ and $B$ depend on it. However, we can transform the summation modes in such a way to transport from one representation to another, meaning that they are all equivalent. 
	
	One can easily notice from Eq.~\eqref{Eq:IG_nearbulk} that the zeroth mode now represents the bulk scenario (if we take $L_\alpha \rightarrow \infty$ we get an exponential supression, so the only survivor is $n_\alpha^{(\ell)}=0$). This will produce the standard representation,
	\begin{equation*}
	\mathcal{I}_G = C_G 
	\left[\prod_i \int_0^\infty du_i\right]
	\frac{e^{-\sum_{i }u_i m_i^2}}{(4\pi)^{\frac{D}{2}L}}
	\frac{e^{-\frac{V(P;u)}{U(u)}}}{U(u)^{\frac{D}{2}}},
	\end{equation*}
	\noindent as should be expected. This shows the consistency of the representation, that recovers the expected scenario when the compactification is removed. As far as we know, this simple and expected connection between the parametric representation in the compactified scenario $(D,d)$ and the non-compactified scenario $(D)$ has never been explicitly stated.
	
	In the following, we change the parametrization and proceed to show that the usual sum over the Bessel-K functions appears naturally. This is expected when dealing with Feynman amplitudes in periodically compactified spaces as shown in previous works~\cite{Khanna:2009zz,Khanna:2014qqa,Elizalde:2012zza} restricted to contributions of Feynman graphs of low orders.
	
	\subsection{Change of parametrization}
	
	We employ a new parametrization that sectorize the $u_i$-parameters such that there is one parameter ($s$) that runs over the positive real axis and all others ($t_i$) just lie in the interval $[0,1]$. This is done by the sectorization
	\begin{align*}
	u_1 &= s t_1 \ldots t_{I-1},\\
	u_2 &= s t_1 \ldots t_{I-2} (1 -t_{I-1}),\\
	u_3 &= s t_1 \ldots t_{I-3} (1-t_{I-2}),\\
	\vdots &= \vdots\nonumber\\
	u_{I-1} &= s t_1(1-t_2),\\
	u_I &= s (1-t_1).
	\end{align*}
	This change of variables produce $\prod_{i=1}^I du_i = s^{I-1} ds \prod_{i=1}^{I-1} t_i^{I-1-i}  dt_i$. And, as all $u_i$ have a factor $s$, this will factor out in a way that allows us to define
	\begin{subequations}
		\begin{align}
		U(u_i) &= s^{L} \overline{U} (t_i),\\
		V(P;u_i) &= s^{L+1} \overline{V} (P;t_i),\\
		A(u_i) &= s^{L-1} \overline{A}(t_i),\\
		\widetilde{B}(u_i) &= s \overline{\widetilde{B}}(t_i).
		\end{align}
	\end{subequations}
	\noindent Notice that one must determine the new expressions by direct inspection.
	
	With this notation, the amplitude related to the diagram becames
	\begin{equation}
	\mathcal{I}_G = C_G 
	\int_0^\infty ds s^{I-1} \left[\prod_{i=1}^{I-1} \int_0^1 dt_i t_i^{I-1-i}\right] 
	\frac{e^{-s\mathcal{M}^2(t_i,m_i)}}{(4\pi)^{\frac{D}{2}L}}
	\frac{e^{-s\frac{\overline V(P;t)}{\overline U(t)}}}{s^{\frac{DL}{2}}{\overline U}^{\frac{D}{2}}} 
	\sum_{ \substack{n_\alpha^{(\ell)}\in \mathbb{Z}\\\forall \ell, \forall \alpha}}
	e^{
		-\frac{1}{s} \sum_\alpha \frac{L_\alpha^2}{4 \overline U(t)} \textbf{n}_\alpha^{t} \overline A(t) \textbf{n}_\alpha
	}
	e^{
		\frac{2\pi i}{\overline U(t)} \sum_{\alpha} \textbf{n}_\alpha^t \overline A(t) \overline{\widetilde{B}}_\alpha
	},
	\end{equation}
	\noindent with
	\begin{equation}
	\mathcal{M}^2[t_i,m_i^2] = \sum_i \frac{u_i}{s} m_i^2.
	\end{equation}
	\noindent When $m_i=m$ we get simply $\mathcal{M}^2 = m^2$.
	
	By inspection, we have two kinds of integrals regarding the parameter $s$. When all modes are zero ($n_{\alpha}^{(\ell)} = 0$) we have a gamma function, when any of them is different from zero we get the integral that defines the modified Bessel function of the second kind (Bessel-K), that is
	\begin{multline}
	\mathcal{I}_G = C_G 
	\left[\prod_{i=1}^{I-1} \int_0^1 dt_i t_i^{I-1-i}\right] 
	\frac{\Gamma\left(I-\frac{DL}{2}\right)}{(4\pi)^{\frac{D}{2}L}{\overline U}^{\frac{D}{2}}}
	\frac{1}{\left[\mathcal{M}^2(t_i,m_i) + \frac{\overline V(P	;t)}{\overline U(t)} \right]^{I-\frac{DL}{2}}} 
	\\
	+ \frac{2 C_G}{(4\pi)^{\frac{D}{2}L}} \left[\prod_{i=1}^{I-1} \int_0^1 dt_i t_i^{I-1-i}\right] 
	\sum^{'}_{ \substack{n_\alpha^{(\ell)}\in \mathbb{Z}\\\forall \ell, \forall \alpha}}
	e^{
		\frac{2\pi i}{\overline U(t)} \sum_{\alpha} \textbf{n}_\alpha^t \overline A(t) \overline{\widetilde{B}}_\alpha
	}
	\left(\frac{\sum_\alpha L_\alpha^2 \textbf{n}_\alpha^{t} \overline A(t) \textbf{n}_\alpha}{4 \overline U(t)\left[\mathcal{M}^2(t_i,m_i) + \frac{\overline V(P;t)}{\overline U(t)} \right]} \right)^{\frac{I-DL/2}{2}} \times\\
	K_{I-\frac{DL}{2}}\left(
	\sqrt{\frac{\sum_\alpha L_\alpha^2 \textbf{n}_\alpha^{t} \overline A(t) \textbf{n}_\alpha}{\overline U(t)} }
	\sqrt{\mathcal{M}^2(t_i,m_i) + \frac{\overline V(P;t)}{\overline U(t)}}
	\right).
	\label{Eq:Amplitude_NearBulk}
	\end{multline}
	\noindent Where $\sum^{'}$ represents that we do not sum at the point where all modes are zero.
	
	This completes the analysis. Once again we remark that this expression is suitable for large $L_\alpha$. This is much more evident at this point. If we try to extract information from small $L_\alpha$ we approach the limit $\lim_{z\rightarrow0}K_\nu(z)$, where the Bessel-K function diverge and one must be very careful with the treatment, as discussed in Ref.~\cite{Cavalcanti:2018pgi,Cavalcanti:2019mli}. Also, as the length parameter, $L_\alpha$ is reduced each term of the sum gets bigger and the convergence of the summations takes longer. This is the justification why we also exhibit the scenario near the dimensional reduction in Sec.~\ref{Sec:NearRedDim}.
	
	In the following section, we show the needed terms $U,V,A$ and $B$ for some Feynman graphs.
	
	%%%%%%%%%%%%%%%%%%%%%%%%%%%%%%%%%%%%%%%%%%%%%%%%%%%%%%%%%%%%%%%
	%%%%%%%%%%%%%%%%%%%%%%%%%%%%%%%%%%%%%%%%%%%%%%%%%%%%%%%%%%%%%%%
	%%%%%%%%%%%%%%%%%%%%%%%%%%%%%%%%%%%%%%%%%%%%%%%%%%%%%%%%%%%%%%%
	
	\section{Some examples in $\phi^3$}\label{Sec:Diagrams}
	
	In this section, we assume that our model is the simple toy model $g\phi^3$ and exhibit the relevant information to write the amplitudes of certain diagrams both in the near dimensional reduction representation, Sec.~\ref{Sec:NearRedDim}, and the near bulk representation, Sec.~\ref{Sec:NearBulk}, for systems living in periodically compactified topologies. The required information, for each diagram, to show the two Symanzik polynomials ($U$ and $V$) and some complementary information as the matrix $M$ (see Eq.~\eqref{Eq:M_def}), $A$ (recall that $A = M^{-1}\det M = U M^{-1}$) and the vector $B$ (see Eq.~\eqref{Eq:B_def}). In what follows one can obtain the Symanzik polynomials by any of the three equivalent expressions Eq.~\eqref{Eq:Symanzik}, Eq.~\eqref{Eq:SymanzikTree} ou Eq.~\eqref{Eq:SymanzikTodorov}.
	
	As a first example we consider the fish diagram, Eqs.~\eqref{Eqs:fish}, with external momenta $Q$. Due to the existence of only one loop, the matrix is reduced to a number. We define for convinience that $u_{12} = u_1+u_2$, and we follow this notation in all other examples.
	\begin{figure}[h]
		\centering\includegraphics[width=3cm]{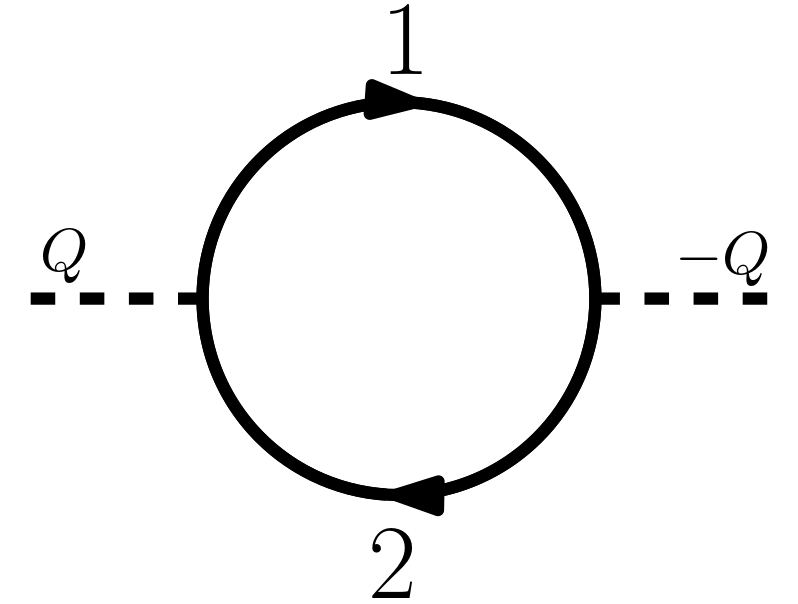}
		\caption{Fish diagram}
		\label{G_Fish}
	\end{figure}
	
	\begin{subequations}
		\label{Eqs:fish}
		\begin{align}
		U &= u_{12} = s\\
		V/Q^2 &= u_1u_2 = s^2 \left\{t(1-t)\right\}\\
		M &= u_{12} = s\\
		A &= 1\\
		B &= u_2 \omega^p_\alpha = s(1-t) \omega^p_\alpha
		\end{align}
	\end{subequations}
	
	In this simple scenario, the expression near bulk is
	
		\begin{multline}
	\mathcal{I}_G = C_G 
	\int_0^1 dt
	\frac{\Gamma\left(2-\frac{D}{2}\right)}{(4\pi)^{\frac{D}{2}}}
	\frac{1}{\left[\mathcal{M}^2(t_i,m_i) + P^2 t(1-t) \right]^{2-\frac{D}{2}}} 
	+ \frac{2 C_G}{(4\pi)^{\frac{D}{2}}} \int_0^1 dt 
	\sum^{'}_{ \substack{n_\alpha\in \mathbb{Z}\\\forall \alpha}}
	e^{
		2\pi i \sum_{\alpha} (1-t)n_\alpha n_\alpha^p
	}
	\times\\
	\left(\frac{\sum_\alpha L_\alpha^2 n_\alpha^2}{4 \left[\mathcal{M}^2(t_i,m_i) + P^2 t(1-t) \right]} \right)^{\frac{2-D/2}{2}} K_{2-\frac{D}{2}}\left(
	\sqrt{\sum_\alpha L_\alpha^2 n_\alpha^2 }
	\sqrt{\mathcal{M}^2(t_i,m_i) + P^2 t(1-t)}
	\right).
	\end{multline}
	
	The first component is the well-known bulk contribution and the second component is the finite-size contribution. We can get a bit further and study, for example, the asymptotic behavior of this amplitude with respect to the external momenta $P$. This falls outside the scope of this work and we shall return to this point in a near future.
	
\begin{figure}[h]
		\centering
		\includegraphics[width=3cm]{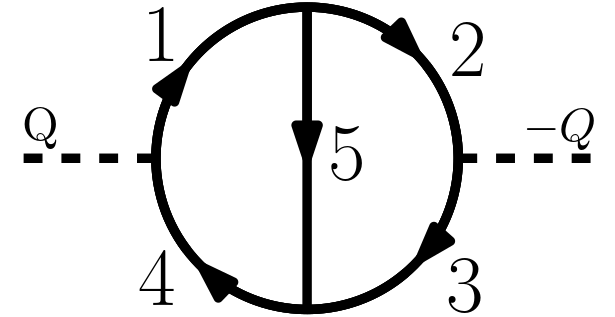}
		\caption{1PI diagrams at two loops}
		\label{G_Ball1}
	\end{figure}

\begin{figure}[h]
	\centering
	\includegraphics[width=3cm]{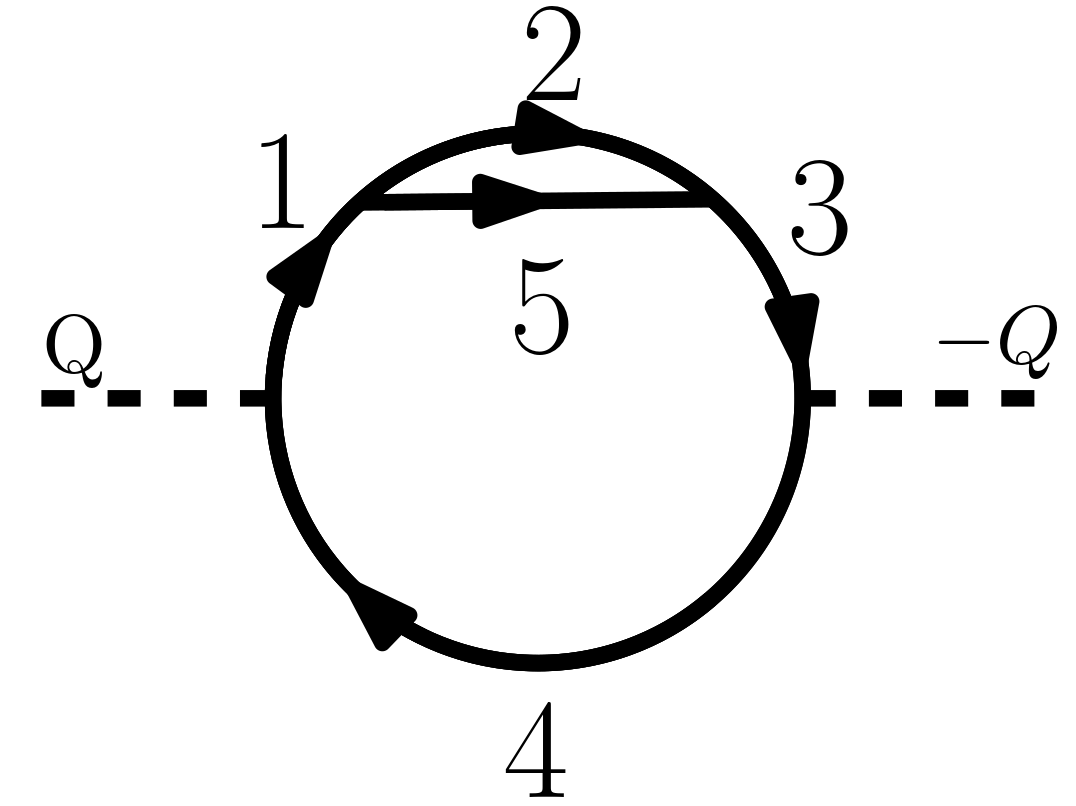}
	\caption{1PI diagrams at two loops}
	\label{G_Ball2}
	\end{figure}

	With regard to mass corrections there are two 1PI (one-particle irreducible) diagrams at two loops (see Fig.~\ref{G_Ball1} and Fig.~\ref{G_Ball2}). For first diagram, Fig.~\ref{G_Ball1}, the functions allows to reproduce both the near dimensional reduction and near bulk scenario are exhibited in Eqs.~\eqref{Eqs:ball1}. 
	
	\begin{subequations}
		\label{Eqs:ball1}
		\begin{align}
		U &= u_5u_{1234}+u_{14}u_{23} = s^2 \left\{ t_1\left[1-t_1+t_1t_2\left(1-t_3t_4-t_2(1-t_3t_4)^2\right)\right] \right\}\\
		V/Q^2 &= u_5u_{12}u_{34} + u_{12}u_3u_4+u_1u_2u_{34} \nonumber\\&= s^3\left\{
		t_1t_2^2t_3\left[1-t_1-t_2t_3+t_1t_2t_3 + t_1t_4(1-t_2)(1-t_3t_4)+t_1^2(1-t_4)(1-t_3)(1-t_2(1-t_3t_4))\right]
		\right\}
		\\
		M &= \begin{pmatrix}
		u_{145}&-u_5\\-u_5&u_{235}	
		\end{pmatrix}
		=
		s \begin{pmatrix}
		1-t_2+t_1t_2t_3t_4&-1+t_1\\-1+t_1&1-t_1+t_2(1-t_3t_4)	
		\end{pmatrix}
		\\
		A &= \begin{pmatrix}
		u_{235}&u_5\\u_5&u_{145}	
		\end{pmatrix}
		=
		s \begin{pmatrix}
		1-t_1+t_2(1-t_3t_4)&1-t_1\\1-t_1&1-t_2+t_1t_2t_3t_4	
		\end{pmatrix}
		\\
		B &= \omega^p_\alpha \begin{pmatrix}
		u_4\\u_3
		\end{pmatrix} 
		= s t_1 \omega^p_\alpha \begin{pmatrix}
		1-t_2\\t_2(1-t_3)
		\end{pmatrix}
		\end{align}
	\end{subequations}
	
	For the second 1PI diagram, see Fig.~\ref{G_Ball2}, the two scenarios are reproduced if we use the functions in Eqs~\eqref{Eqs:ball2}.	
	\begin{subequations}
		\label{Eqs:ball2}
		\begin{align}
		U &= u_5u_{2}+u_{134}u_{25} = s^2t_1\left\{
		\left[1-t_2t_3(1-t_4)\right]\left[1-t_1+t_1t_2t_3(1-t_4)\right] + t_2t_3(1-t_1)(1-t_4)
		\right\}\\
		V/Q^2 &= u_4u_{2}u_{5} + u_{4}u_{13}u_{25}\nonumber \\ 
		&= s^3t_1^2t_2\left\{
		t_3(1-t_1)(1-t_2)(1-t_4)+(1-t_2)(1-t_3+t_3t_4)(1-t_1+t_1t_2t_3(1-t_4))
		\right\}\\
		M &= \begin{pmatrix}
		u_{1345}&-u_5\\-u_5&u_{25}	
		\end{pmatrix}
		= 
		s \begin{pmatrix}
		1-t_1t_2t_3+t_1t_2t_3t_4&-1+t_1\\-1+t_1&1-t_1+t_1t_2t_3(1-t_4)	
		\end{pmatrix}
		\\
		A &= \begin{pmatrix}
		u_{25}&u_5\\u_5&u_{1345}	
		\end{pmatrix} = 
		s \begin{pmatrix}
		1-t_1+t_1t_2t_3(1-t_4)&1-t_1\\1-t_1&1-t_1t_2t_3+t_1t_2t_3t_4	
		\end{pmatrix}\\
		B &= \omega^p_\alpha \begin{pmatrix}
		u_4\\0
		\end{pmatrix} = st_1\omega^p_\alpha \begin{pmatrix}
		1-t_2\\0
		\end{pmatrix}
		\end{align}
	\end{subequations}	
	
\begin{figure}[h]
	\centering
	\includegraphics[width=6cm]{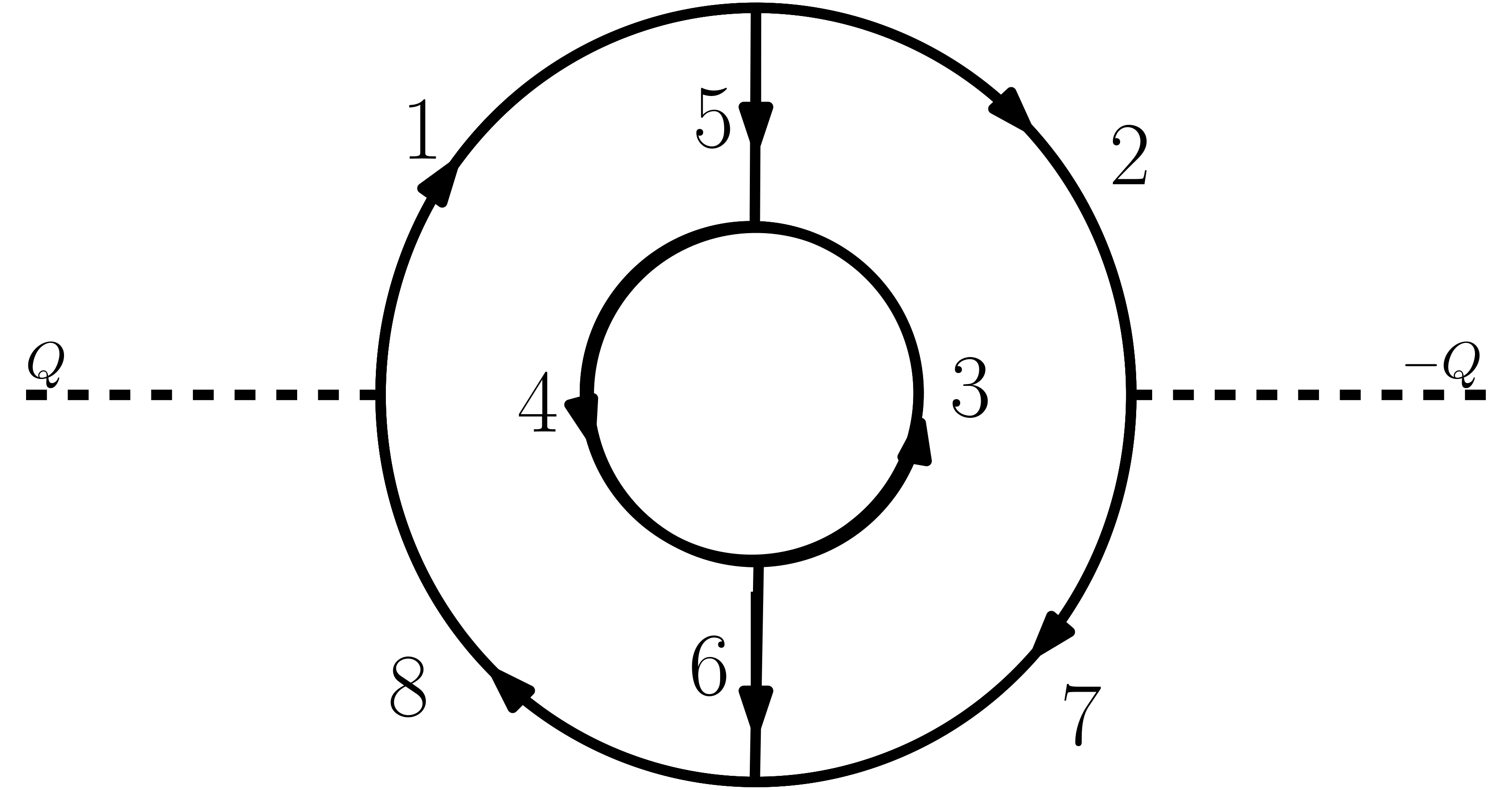}
	\caption{Example with three loops}
	\label{G_Pokeball}
\end{figure}

	At three loops we have a large number of 1PI diagrams that contribute to the correction of the full propagator. One such example is shown in Fig.~\ref{G_Pokeball}. One can use directly the matrices $A$ and the vector $B$ to build the representation near the bulk, see Eq.~\eqref{Eq:Amplitude_NearBulk}, and use $M$ to build the representtion near the dimensional reduction, see Eq.~\eqref{Eq:Amplitude_NearRedDim}.
	
\begin{subequations}
	\label{Eqs:pokeball}
	\begin{align}
	U = u_{34}u_{56}u_{1278}+u_{34}u_{18}u_{27}
	+u_3u_4u_{1278}\\
	V/Q^2 = u_{12}u_{34}u_{56}u_{78} + u_1u_2u_{34}u_{78}\nonumber\\+u_{12}u_3u_4u_{78}+u_{12}u_{34}u_7u_8\\
	M = \begin{pmatrix}
	u_{14568}&-u_{456}&u_4\\
	-u_{456}&u_{24567}&-u_4\\
	u_4& -u_4 & u_{34}
	\end{pmatrix}\\
	A = \begin{pmatrix}
	\begin{matrix}
	u_{2567}u_{34}\\+u_3u_4
	\end{matrix}&\begin{matrix}
	u_3u_4\\+u_{34}u_{56}
	\end{matrix}&-u_4u_{27}\\
	\begin{matrix}
	u_3u_4\\+u_{34}u_{56}
	\end{matrix}&\begin{matrix}
	u_{1568}u_{34}\\+u_3u_4
	\end{matrix}&u_4u_{18}\\	-u_4u_{27} & u_4u_{18} & \begin{matrix}
	u_{456}u_{1278}\\+u_{18}u_{27}
	\end{matrix}
	\end{pmatrix}\\
	B = \omega^p_\alpha \begin{pmatrix}
	u_8\\u_7\\ 0
	\end{pmatrix}
	\end{align}
\end{subequations}

	\section{Conclusion}
	
	We manage to show two useful parametric representations for Feynman diagrams in periodically compactified spaces. One representation is more suitable near the bulk and recovers the expected standard Schwinger parametric representation for non-compactified theories at the bulk. The other representation is more suitable near the dimensional reduction and recovers that only a completely static mode approximation can generate a dimensional reduction. Both expressions allow one to obtain the behavior with respect to the compactification length and shall be helpful for those interested in higher-order Feynman diagrams in compactified spaces to study a myriad of phenomena as depicted by works in quantum field theory with compactifications~\cite{Landsman:1989be,Khanna:2009zz,Khanna:2014qqa,Bordag:2001qi,Milton:2004ya,Klimchitskaya:2009cw,Elizalde:2012zza,Mogliacci:2018oea}. 
	
	Of course, we still do not get here the full picture and there are many remaining tasks to be done: 1 - There is a need to extend the present work for fields with non-zero spins; 2 - It might be useful to consider scenarios where space is compactified by a different prescription other than a periodic one (Dirichlet, Neumann, Robin, ...); 3 - It is of utmost importance to extract asymptotic information from the parametric representations so we can easily understand the behavior of a higher-order graph in compactified space. Each of these topics are the subject of future works. 
	
	\acknowledgments{The author thanks the Brazilian agency Conselho Nacional de Desenvolvimento Cient\'ifico e Tecnol\'ogico (CNPq) for financial support.}
	
	\appendix

	\bibliography{0AllRefs}{}

\end{document}